\documentclass[conference]{IEEEtran}
\IEEEoverridecommandlockouts

\usepackage{cite}
\usepackage{amsmath,amssymb,amsfonts}
\usepackage{algorithmic}
\usepackage{graphicx}
\usepackage{textcomp}
\usepackage{xcolor}
\usepackage{algorithm}
\usepackage{algorithmic}
\usepackage{url}
\usepackage{bm}
\usepackage{siunitx}
\usepackage{booktabs,multirow,makecell,threeparttable}
\usepackage[caption=false,font=footnotesize]{subfig}

\usepackage{tikz}
\usetikzlibrary{positioning}
\usepackage{tkz-graph}
\usepackage{pgfplots}
\usepackage{listofitems}

\usetikzlibrary{math, calc, positioning, shapes, backgrounds, fit, arrows, decorations.pathreplacing}
\pgfplotsset{compat=1.16}

\def\BibTeX{{\rm B\kern-.05em{\sc i\kern-.025em b}\kern-.08em
    T\kern-.1667em\lower.7ex\hbox{E}\kern-.125emX}}
\begin{document}

\title{Scalable Mamba-Based Message-Passing Neural Decoder for Error-Correcting Codes}

\author{
  \IEEEauthorblockN{Rostislav Gusev, Nikita Aleksandrov, Artem Solomkin, Dmitry Artemasov}

  \IEEEauthorblockA{
  \textit{Skolkovo Institute of Science and Technology}, 
  Moscow, Russia \\
  \{rostislav.gusev, nikita.aleksandrov, artem.solomkin, d.artemasov\}@skoltech.ru}
}

\maketitle

\begin{abstract}
Forward error correction is essential for reliable communication over noisy channels. Attention-based model-free neural decoders have shown strong performance for short codes, but their scalability to longer codes is limited by the quadratic memory and computational cost of attention. In this paper, we introduce the Mamba message-passing decoder (MMPD), an attention-matrix-free syndrome-based neural decoder for binary linear codes. MMPD retains the Tanner-graph structure of a message-passing decoder by performing local pairwise aggregation along variable-check edges. To enable efficient long-range information propagation, these local updates are combined with bidirectional Mamba state-space blocks. By avoiding dense attention matrices, MMPD scales more favorably for long codes in both memory and computation. Experiments on the $\bm{(1056, 880)}$ LDPC code show that MMPD achieves a $\bm{0.45}$~dB gain over the state-of-the-art CrossMPT decoder at a specified target bit error rate, while reducing memory consumption by a factor of $\bm{1.5}$. This reduction factor increases substantially for longer codes, demonstrating the applicability of MMPD to scalable neural decoding of practical long codes.
\end{abstract}

\begin{IEEEkeywords}
Attention-matrix-free decoder, binary linear codes, deep neural networks, forward error correction, Mamba, message passing, model-free neural decoder.
\end{IEEEkeywords}

\section{Introduction}
\label{sec:intro}

Forward error correction is a fundamental component of modern communication systems, enabling reliable transmission over noisy channels under strict latency and complexity constraints. Neural decoders have attracted growing interest as a way to narrow the gap between computationally tractable classical decoding algorithms and the performance of optimal but intractable decoding rules, especially in finite-length regimes~\cite{Matsumine2024}. Within the development of neural decoding techniques, two main directions have emerged: model-based methods, which unfold and augment classical decoders such as belief propagation (BP) with trainable parameters~\cite{Cammerer2022, Andreev2021}, and model-free methods, which learn the decoding rule in a data-driven manner. Among the latter, syndrome-based decoding has become one of the most promising approaches for binary linear codes, because code linearity allows training the model on the all-zero codeword without overfitting and recasts decoding as the recovery of the channel-induced error pattern~\cite{Bennatan2018, Artemasov2024}.

A significant step in neural decoding was the introduction of the error correction code Transformer (ECCT)~\cite{Choukroun2022ECCT}, a Transformer-based architecture for decoding linear codes. ECCT demonstrated that masked attention can model dependencies between code bits more effectively than prior neural decoding architectures. Subsequent works showed that Transformer decoders can be further improved by injecting the Tanner-graph structure more explicitly. In particular, in CrossMPT~\cite{Park2025}, decoding is reformulated as alternating message passing between magnitude and syndrome streams, using parity-check-matrix-induced masks inside cross-attention so that variable and check representations exchange information only along valid graph connections. This message-passing interpretation substantially improved the structured reasoning capabilities of Transformer decoders and established a strong baseline for neural decoding of short Bose-Chaudhuri-Hocquenghem (BCH), Polar, and low-density parity-check (LDPC) codes.

Recent work has broadened neural ECC decoding beyond a single architectural family. The accelerating error correction code Transformer (AECCT)~\cite{Levy2024} extends Transformer-based decoding through ternary quantization and code-aware multi-head processing.  In a complementary direction, the hybrid Mamba–Transformer decoder (ECCM)~\cite{Cohen2025} combines Mamba blocks with Transformer layers to improve the speed–accuracy trade-off while retaining attention. In parallel, studies~\cite{Choukroun2024} and~\cite{Park2025CodeAgnostic} explored the development of code-agnostic models aimed at improving generalization and reuse across multiple code families. Another important direction treats decoding as a denoising problem rather than a purely attention-based prediction task. Paper~\cite{Choukroun2022DDECC} introduced a diffusion-based decoder (DDECC) with iterative reverse denoising guided by parity information, whereas a subsequent study~\cite{Lei2026} reduces this cost by moving to a one-step consistency-style decoder (ECCFM) with lower latency and improved error-rate performance. Together, these directions show that the field is moving beyond raw decoding accuracy toward neural decoders that are more efficient, scalable, and reusable.

Despite this progress, an important gap remains. The strongest transformer-based decoders still rely on attention as the main interaction primitive, which makes memory and runtime scale poorly as block length grows, even when masking is used to enforce code structure. Efficiency-oriented variants alleviate this issue, but they remain fundamentally attention-based. As a result, the field lacks a decoder that retains the structured variable-check reasoning of modern message-passing transformers while replacing dense attention with graph interactions whose cost scales with the number of Tanner-graph edges.

In this work, we address this gap by introducing the Mamba message-passing decoder (MMPD), which preserves the core inductive bias of Tanner-graph decoding while avoiding dense attention matrices in the main update path. Following the syndrome-based formulation, the decoder maintains separate variable-node and check-node streams and alternates information exchange across the two streams, as in CrossMPT. However, instead of Transformer-style masked cross-attention, local interactions are implemented through compact pairwise aggregation along Tanner-graph edges, followed by gated residual node updates and bidirectional Mamba blocks that provide efficient global propagation across the variable and check streams.

The proposed method differs from prior work in several key respects. Relative to ECCT and CrossMPT, we replace Transformer-style cross-attention with edge-restricted pairwise scoring combined with state-space global mixing, avoiding dense attention matrices. Compared with AECCT, MMPD changes the decoder backbone itself and its scaling behavior rather than compressing a Transformer architecture. Unlike the hybrid Mamba--Transformer design of ECCM, MMPD uses state-space modeling for global mixing on top of graph-structured local aggregation. Relative to DDECC and ECCFM, MMPD follows a direct discriminative decoding formulation rather than a generative denoising approach.

The main contributions of this paper are threefold. First, we introduce an attention-matrix-free neural decoder that combines edge-restricted pairwise message aggregation with bidirectional state-space global mixing, while preserving the structured variable-check decomposition of modern message-passing transformers. Second, we provide a systematic comparison with existing neural decoders, demonstrating that MMPD improves the performance-memory trade-off relative to state-of-the-art attention-based baselines. Third, we target the practically important regime of long codes, showing that the reduced memory consumption of the proposed design enables scaling to code lengths for which existing neural decoders become infeasible.

\section{System Model}
\label{sec:system-model}
We consider a communication system in which a transmitter sends a $k$-bit information word $\mathbf{u} \in \{0,1\}^k$. The information word is encoded into a codeword $\mathbf{c}\in\{0,1\}^n$ using a binary linear block code $\mathcal{C}$ of length $n$ and dimension $k$. The codeword is then mapped to a binary phase-shift keying (BPSK) signal,
\begin{equation}
    \mathbf{x} = \tau(\mathbf{c}), \quad \tau(\mathbf{c}) = (\tau(c_{1}), \ldots, \tau(c_{n})),
\end{equation}
where $\tau:\{0, 1\} \rightarrow \{1, -1\}$.

The modulated codeword $\mathbf{x}\in\{-1,1\}^n$ is transmitted over an additive white Gaussian noise (AWGN) channel. The received vector is
\begin{equation}
\label{eq:awgn}
\mathbf{y} = \mathbf{x} + \mathbf{z},
\end{equation}
where $\mathbf{y} = (y_1, \ldots, y_n) \in \mathbb{R}^{n}$, $\mathbf{z}\sim \mathcal{N}(0,\sigma^2 \mathbf{I}_n)$, and $\mathbf{I}_n$ denotes the $n \times n$ identity matrix.

For the graph-based representation of the code, we consider the Tanner graph associated with the parity-check matrix $\mathbf{H}\in\{0,1\}^{n-k\times n}$. The graph consists of $n$ variable nodes (VNs) and $n-k$ check nodes (CNs). A check node $j$ and a variable node $i$ are connected if $\mathbf{H}_{j,i}=1$. In the following, $\mathcal{N}_v(i)$ denotes the set of check nodes connected to variable node $i$, and $\mathcal{N}_c(j)$ denotes the set of variable nodes connected to check node $j$.

\section{Mamba-Based Message-Passing Decoding}
\label{sec:architecture}

In this work, we use the syndrome-based formulation introduced in~\cite{Bennatan2018}. The decoder takes as input the reliability vector $\mathbf{m}_y = |\mathbf{y}| \in \mathbb{R}^n$ and the signed syndrome vector $\mathbf{s}_y \in \{-1,+1\}^{n-k}$. The latter is computed as $\tau(\mathbf{H}\mathbf{y}_b)$, where $\mathbf{y}_b = \operatorname{bin}(\operatorname{sign}(\mathbf{y}))$ denotes the demodulated version of $\mathbf{y}$.

\begin{figure*}
    \centering
    \includegraphics[width=0.95\textwidth, trim={0 1.5cm 0 0.5cm}]{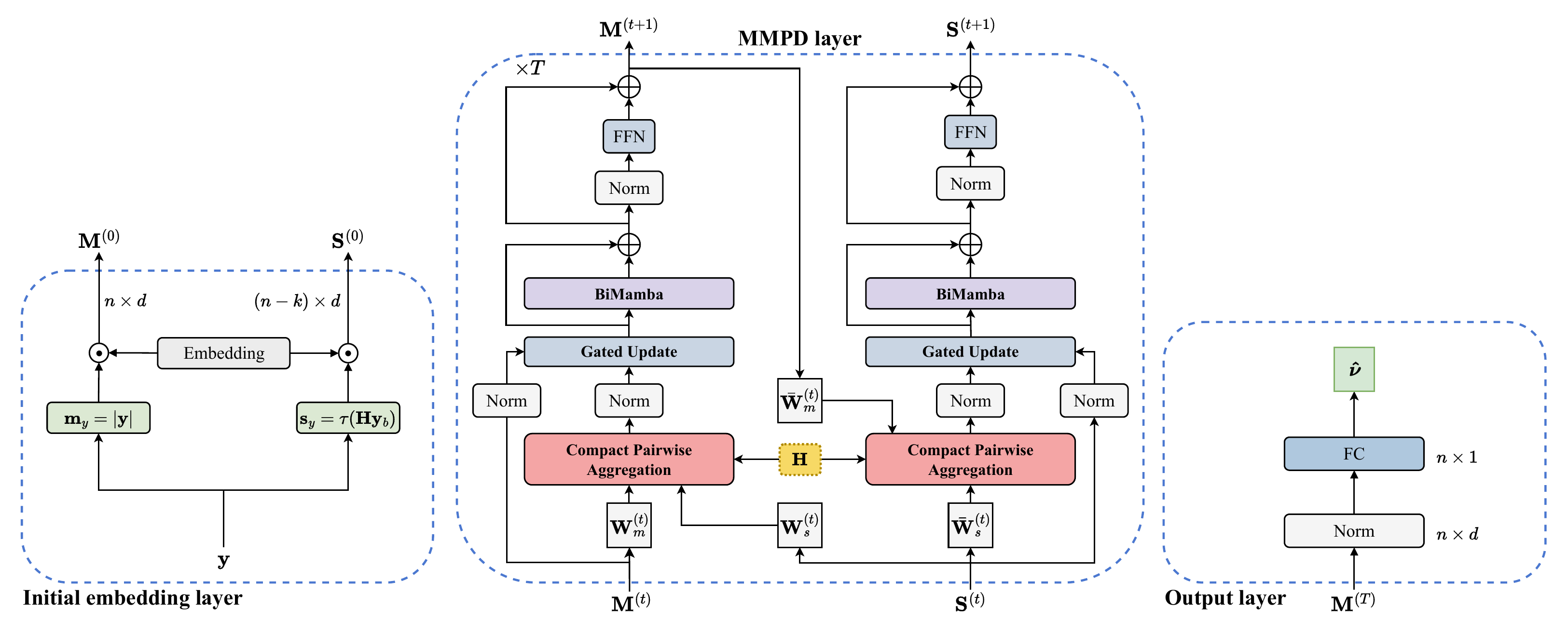}
    \caption{Architecture of the proposed MMPD.}
    \label{fig:decoder-architecture}
    \vspace{-10pt}
\end{figure*}

Based on the neural message-passing approach of~\cite{Park2025}, the proposed decoder maintains two latent streams throughout the decoding process: a VN stream and a CN stream. At decoding block $t \in \{0,\dots,T-1\}$, these streams are represented by
\begin{align}
\mathbf{M}^{(t)} &= \left[\mathbf{m}^{(t)}_{1},\dots,\mathbf{m}^{(t)}_{n}\right]^\top \in \mathbb{R}^{n \times d},\\
\mathbf{S}^{(t)} &= \left[\mathbf{s}^{(t)}_{1},\dots,\mathbf{s}^{(t)}_{n-k}\right]^\top \in \mathbb{R}^{n-k  \times d},
\end{align}
where $d$ denotes the model's hidden dimension, and $(\cdot)^\top$ denotes the transpose operation. \figurename~\ref{fig:decoder-architecture} provides an overview of the proposed decoding pipeline, including input initialization, iterative VN–CN message passing, bidirectional Mamba refinement, and final prediction by the output head.

\subsubsection{Input initialization}

The initial node states are obtained by scaling learnable node-specific embeddings by the corresponding scalar inputs
\begin{equation}
\mathbf{m}_i^{(0)} = {m_y}_i \mathbf{e}^{(v)}_{i} + \mathbf{b}^{(v)}_{i}, \qquad i=1,\dots,n,
\label{eq:emb-mag}
\end{equation}
\begin{equation}
\mathbf{s}_j^{(0)} = {s_y}_j \mathbf{e}^{(c)}_{j} + \mathbf{b}^{(c)}_{j}, \qquad j=1,\dots,n-k,
\label{eq:emb-synd}
\end{equation}
where $\mathbf{e}^{(v)}_{i}, \mathbf{b}^{(v)}_{i} \in \mathbb{R}^{d}$ and $\mathbf{e}^{(c)}_{j}, \mathbf{b}^{(c)}_{j} \in \mathbb{R}^{d}$ are learnable VN and CN embeddings, respectively.

\subsubsection{Compact pairwise aggregation}

To aggregate information from CNs to VNs, we first project the node states into an $r$-dimensional space
\begin{equation}
\tilde{\mathbf{m}}_i^{(t)} = \mathbf{W}_m^{(t)} \mathbf{m}_i^{(t)}, \qquad
\tilde{\mathbf{s}}_j^{(t)} = \mathbf{W}_s^{(t)} \mathbf{s}_j^{(t)},
\end{equation}
where $\mathbf{W}_m^{(t)}, \mathbf{W}_s^{(t)} \in \mathbb{R}^{r \times d}$ are trainable matrices.

For each Tanner-graph edge $(j,i)$ such that $\mathbf{H}_{j,i}=1$, we form a compact pairwise feature~\cite{Mou2016}
\begin{equation}
\boldsymbol{\phi}_{ij}^{(t)} =
\left[
\tilde{\mathbf{m}}_i^{(t)} \odot \tilde{\mathbf{s}}_j^{(t)}
\ ;\
\tilde{\mathbf{m}}_i^{(t)} - \tilde{\mathbf{s}}_j^{(t)}
\right]
\in \mathbb{R}^{2r},
\end{equation}
and compute a scalar edge score
\begin{equation}
f_{ij}^{(t)} =
\mathbf{w}_2^{(t)\top}
\operatorname{SiLU}\!\left(\mathbf{W}_1^{(t)} \boldsymbol{\phi}_{ij}^{(t)}\right),
\end{equation}
where $\mathbf{W}_1^{(t)} \in \mathbb{R}^{r \times 2r}$ and $\mathbf{w}_2^{(t)} \in \mathbb{R}^{r}$ are learnable parameters, and $\operatorname{SiLU}$ denotes the sigmoid linear unit activation.

The scores are then normalized over the neighbors of VN $i$
\begin{equation}
\alpha_{ij}^{(t)} = \frac{\exp(f_{ij}^{(t)})}
{\sum\limits_{j' \in \mathcal{N}_v(i)} \exp(f_{ij'}^{(t)})},
\qquad j \in \mathcal{N}_v(i).
\end{equation}
The resulting CN-to-VN message is
\begin{equation}
\mathbf{r}_i^{(t,v)} =
\mathbf{W}_o^{(t,v)}
\sum_{j \in \mathcal{N}_v(i)}
\alpha_{ij}^{(t)} \, \mathbf{W}_p^{(t,v)} \mathbf{s}_j^{(t)},
\end{equation}
where $\mathbf{W}_p^{(t,v)}, \mathbf{W}_o^{(t,v)} \in \mathbb{R}^{d \times d}$ are learnable VN-stream projections.

The reverse VN-to-CN aggregation is defined similarly.
\begin{equation}
\bar{\mathbf{s}}_j^{(t)} = \bar{\mathbf{W}}_s^{(t)} \mathbf{s}_j^{(t)}, \qquad
\bar{\mathbf{m}}_i^{(t+1)} = \bar{\mathbf{W}}_m^{(t)} \mathbf{m}_i^{(t+1)},
\end{equation}
\begin{equation}
\bar{\boldsymbol{\phi}}_{ji}^{(t)} =
\left[
\bar{\mathbf{s}}_j^{(t)} \odot \bar{\mathbf{m}}_i^{(t+1)}
\ ;\
\bar{\mathbf{s}}_j^{(t)} - \bar{\mathbf{m}}_i^{(t+1)}
\right],
\end{equation}
\begin{equation}
\bar{f}_{ji}^{(t)} =
\bar{\mathbf{w}}_2^{(t)\top}
\operatorname{SiLU}\!\left(\bar{\mathbf{W}}_1^{(t)} \bar{\boldsymbol{\phi}}_{ji}^{(t)}\right),
\end{equation}
\begin{equation}
\bar{\alpha}_{ji}^{(t)} =
\frac{\exp(\bar{f}_{ji}^{(t)})}
{\sum\limits_{i' \in \mathcal{N}_c(j)} \exp(\bar{f}_{ji'}^{(t)})},
\qquad i \in \mathcal{N}_c(j),
\end{equation}
and the VN-to-CN message is
\begin{equation}
\bar{\mathbf{r}}_j^{(t,c)} =
\bar{\mathbf{W}}_o^{(t,c)}
\sum_{i \in \mathcal{N}_c(j)}
\bar{\alpha}_{ji}^{(t)} \, \bar{\mathbf{W}}_p^{(t,c)} \bar{\mathbf{m}}_i^{(t+1)}.
\end{equation}

For brevity, we denote this operation by $\operatorname{Agg}(\cdot)$. Pairwise features are computed with batched GPU tensor operations for indexing, scoring, masked softmax, and reduction, without an edge-wise loop.

\subsubsection{Gated node update}

Both node streams are updated using the same gated residual operator.
Given a current state $\boldsymbol{\xi}$ and an aggregated message $\boldsymbol{\eta}$, the update is computed as
\begin{equation}
\mathbf{h} = \operatorname{GELU}\!\left( \mathbf{W}_h \left[\operatorname{LN}(\boldsymbol{\xi})
\ ;\ \operatorname{LN}(\boldsymbol{\eta})\right]\right),
\end{equation}
\begin{equation}
\mathbf{g} = \sigma(\mathbf{W}_g \mathbf{h}), \qquad
\boldsymbol{\Delta} = \mathbf{W}_{\Delta} \mathbf{h},
\end{equation}
\begin{equation}
\boldsymbol{\xi}' = \boldsymbol{\xi} + \mathbf{g} \odot \boldsymbol{\Delta},
\end{equation}
followed by a feed-forward refinement
\begin{equation}
\operatorname{GUpdate}(\boldsymbol{\xi},\boldsymbol{\eta})
= \boldsymbol{\xi}' + \operatorname{FFN}\!\left(\operatorname{LN}(\boldsymbol{\xi}')\right).
\end{equation}
Here, $\sigma$ denotes the sigmoid function, $\operatorname{LN}$ denotes layer normalization, $\operatorname{GELU}$ represents Gaussian error linear unit, and $\operatorname{FFN}$ denotes the feed-forward network.

Accordingly, the VN update is
\begin{equation}
\hat{\mathbf{m}}_i^{(t+1)} = \operatorname{GUpdate}_v\!\left(\mathbf{m}_i^{(t)}, \mathbf{r}_i^{(t,v)}\right),
\end{equation}
while the CN update is
\begin{equation}
\hat{\mathbf{s}}_j^{(t+1)} = \operatorname{GUpdate}_c\!\left(\mathbf{s}_j^{(t)}, \bar{\mathbf{r}}_j^{(t,c)}\right).
\end{equation}

\subsubsection{Bidirectional Mamba refinement}

After each local aggregation and gated update, a bidirectional Mamba block is applied to the corresponding stream. For an input sequence $\mathbf{X}$, the bidirectional Mamba output is defined as
\begin{equation}
\operatorname{BiMamba}(\mathbf{X})
=
\frac{1}{2}
\left(
\mathcal{M}_{f}(\mathbf{X})
+
\operatorname{rev}\!\left[
\mathcal{M}_{b}(\operatorname{rev}(\mathbf{X}))
\right]
\right),
\end{equation}
where $\mathcal{M}_{f}$ and $\mathcal{M}_{b}$ are independently parameterized Mamba modules~\cite{gu2023mamba} and $\operatorname{rev}(\cdot)$ reverses the sequence order.

\begin{equation}
\bm{\Gamma}^{(t+1)}
=
\text{BiMambaBlock}\!\left(\hat{\bm{\Gamma}}^{(t+1)}\right),
\qquad
\bm{\Gamma}\in\{\mathbf{M},\mathbf{S}\},
\end{equation}
with
\begin{equation}
\begin{aligned}
\text{BiMambaBlock}(\hat{\bm{\Gamma}})
&=
\hat{\bm{\Gamma}}
+ \lambda\,\operatorname{BiMamba}(\hat{\bm{\Gamma}}) \\
&+ \operatorname{FFN}\!\left(
    \operatorname{LN}\!\left(
        \hat{\bm{\Gamma}}
        + \lambda\,\operatorname{BiMamba}(\hat{\bm{\Gamma}})
    \right)
\right),
\end{aligned}
\end{equation}
where $\lambda$ is a learnable scalar.

\begin{table*}[!t]
\caption{Comparison of decoding performance at three SNR values (4, 5, 6) for BP, ECCT~\cite{Choukroun2022ECCT}, AECCT~\cite{Levy2024}, CrossMPT~\cite{Park2025}, and MMPD. The results are measured as the negative natural logarithm of BER ($-\ln(\mathrm{BER})$). The best results are shown in bold, and the second-best results are underlined. Higher is better}\vspace{-20pt}
\label{tab:results}
\begin{center}
\begin{small}
\resizebox{\textwidth}{!}{
\begin{tabular}{ccccccccccccccccc}
\toprule
\multirow{2}{*}{\textbf{Codes}} & \multirow{2}{*}{\textbf{$(n,k)$}}
  & \multicolumn{3}{c}{\textbf{BP} (50 iter.)}
  & \multicolumn{3}{c}{\textbf{ECCT $1.2M$}~\cite{Choukroun2022ECCT}}
  & \multicolumn{3}{c}{\textbf{AECCT $1.2M$}~\cite{Levy2024}}
  & \multicolumn{3}{c}{\textbf{CrossMPT $1.2M$}~\cite{Park2025}}
  & \multicolumn{3}{c}{\textbf{MMPD $1.2M$ (ours)}} \\
\cmidrule(lr){3-5}\cmidrule(lr){6-8}\cmidrule(lr){9-11}\cmidrule(lr){12-14}\cmidrule(lr){15-17}
 & & 4 & 5 & 6
   & 4 & 5 & 6
   & 4 & 5 & 6
   & 4 & 5 & 6
   & 4 & 5 & 6 \\
\midrule
\multirow{3}{*}{Polar}
 & (64,48)  & \makecell{4.26} & \makecell{5.38} & \makecell{6.50}
            & 6.36 & 8.46 & 11.09
            & 6.43 & 8.54 & 11.12
            & \underline{6.51} & \underline{8.70} & \underline{11.31}
            & \textbf{{6.54}} & \textbf{{8.75}} & \textbf{{11.39}} \\
 & (128,86) & \makecell{4.49} & \makecell{5.65} & \makecell{6.97}
            & 6.31 & 9.01 & 12.45
            & 6.04 & 8.56 & 11.81
            & \underline{7.51} & \underline{10.83} & \underline{15.24}
            & \textbf{{7.94}} & \textbf{{11.50}} & \textbf{{16.04}} \\
 & (128,96) & \makecell{4.61} & \makecell{5.79} & \makecell{7.08}
            & 6.31 & 9.12 & 12.47
            & 6.11 & 8.81 & 12.15
            & \underline{7.15} & \underline{10.15} & \underline{13.13}
            & \textbf{7.37} & \textbf{10.33} & \textbf{13.21} \\
\midrule
\multirow{3}{*}{LDPC}
 & (49,24)  & \makecell{6.23} & \makecell{8.19} & \makecell{11.72}
            & 5.79 & 8.13 & 11.40
            & 6.10 & 8.65 & 12.34
            & \underline{6.68} & \underline{9.52} & \underline{13.19}
            & \textbf{6.78} & \textbf{9.67} & \textbf{13.68} \\
 & (121,60) & \makecell{5.65}    & \makecell{8.82}    & \makecell{13.28}
            & 5.01 & 7.99 & 12.78
            & 5.17 & 8.32 & 13.40
            & \underline{5.74} & \underline{9.26} & \underline{14.78}
            & \textbf{5.84} & \textbf{9.62} & \textbf{15.15} \\
 & (121,80) & \makecell{7.84}    & \makecell{11.92}    & \makecell{16.97}
            & 7.07 & 10.96 & 16.25
            & 7.27 & 11.50 & 16.90
            & \underline{7.99} & \underline{12.75} & \underline{18.15}
            & \textbf{8.07} & \textbf{12.83} & \textbf{18.17} \\
\bottomrule
\end{tabular}
}
\end{small}
\end{center}
\vspace{-16pt}
\end{table*}

\subsubsection{Decoder block and output head}

One decoder block consists of the following operations
\begin{align}
\mathbf{R}^{(t,v)} &= \operatorname{Agg}_{c \rightarrow v}(\mathbf{M}^{(t)}, \mathbf{S}^{(t)}; \mathbf{H}), \\
\hat{\mathbf{M}}^{(t+1)} &= \operatorname{GUpdate}_v(\mathbf{M}^{(t)}, \mathbf{R}^{(t,v)}), \\
\mathbf{M}^{(t+1)} &= \operatorname{BiMambaBlock}_v(\hat{\mathbf{M}}^{(t+1)}), \\
\bar{\mathbf{R}}^{(t,c)} &= \operatorname{Agg}_{v \rightarrow c}(\mathbf{S}^{(t)}, \mathbf{M}^{(t+1)}; \mathbf{H}), \\
\hat{\mathbf{S}}^{(t+1)} &= \operatorname{GUpdate}_c(\mathbf{S}^{(t)}, \bar{\mathbf{R}}^{(t,c)}),\\
\mathbf{S}^{(t+1)} &= \operatorname{BiMambaBlock}_c(\hat{\mathbf{S}}^{(t+1)}).
\end{align}
Stacking $T$ such blocks forms the decoder model. The output is produced from the final VN stream by applying the fully connected (FC) layer
\begin{equation}
\hat{\nu}_i = \mathbf{w}_{\mathrm{out}}^\top \operatorname{LN}(\mathbf{m}_i^{(T)}) + b_{\mathrm{out}},
\qquad i=1,\dots,n.
\end{equation}

\subsubsection{Training objective}

Following common practice in the neural decoding literature, the model is trained with the binary cross-entropy loss to predict the error indicator.
\begin{equation}
\mathcal{L} = -\sum_{i=1}^{n} \left[
\varepsilon_i \log \sigma(\hat{\nu}_i)
+ (1-\varepsilon_i)\log\!\bigl(1-\sigma(\hat{\nu}_i)\bigr)
\right],
\end{equation}
where $\boldsymbol{\varepsilon} \in \{0,1\}^n$ is the ground-truth error indicator vector.

\section{Simulation Results}
\label{sec:results}

In this section, we compare the proposed MMPD with classical and state-of-the-art neural decoders:
\begin{itemize}
    \item \textbf{BP}: sum-product belief propagation~\cite{Kschischang2001};
    \item \textbf{ECCT}: error correction code Transformer~\cite{Choukroun2022ECCT};
    \item \textbf{AECCT}: accelerated and compressed version of ECCT~\cite{Levy2024};
    \item \textbf{CrossMPT}: structured cross-attention message-passing Transformer decoder~\cite{Park2025};
    \item \textbf{MMPD}: the proposed Mamba message-passing decoder.
\end{itemize}

For evaluation, random codewords are generated, mapped to BPSK symbols, and transmitted over an AWGN channel as defined in~\eqref{eq:awgn}. Decoding performance is measured using the bit error rate (BER),
\begin{equation}
    \mathrm{BER} = \frac{1}{n}\sum_{i=1}^{n}\mathbb{P}(c_i \neq \hat{c}_i),
\end{equation}
where $\hat{\mathbf{c}}$ denotes the estimated codeword.

We consider two groups of experiments. The first group addresses short-length baseline codes commonly used to benchmark neural decoders, for which performance is reported in terms of $-\ln(\mathrm{BER})$, following common practice in the neural decoding literature~\cite{Choukroun2022ECCT, Levy2024, Park2025}. The second group considers longer codes and showcases the scalability of the developed model, for which performance is presented as BER versus signal-to-noise ratio (SNR) curves. All neural decoders are trained under the same protocol, following~\cite{Park2025}. For the MMPD configuration, we use $T=7$, $d=64$, pairwise rank $r=32$, Mamba state dimension $28$, convolution width $4$, expansion factor $1$, FFN multiplier $1$, and a gated-update hidden dimension of $d/2$. The model is trained for 1000 epochs with AdamW, learning rate $10^{-4}$, weight decay $10^{-4}$, batch size 128, and $E_b/N_0\in[3,7]$~dB.

Table~\ref{tab:results} summarizes the performance results for BP, ECCT, AECCT, CrossMPT, and MMPD on the benchmark codes.\footnote{ECCM~\cite{Cohen2025} is excluded from the performance comparison because the available implementation did not reproduce the results reported in the preprint.}
The $-\ln(\mathrm{BER})$ metric is reported at $E_b/N_0$ values of $4$, $5$, and $6\,\mathrm{dB}$. For this experiment, the hyperparameters of all neural decoders are aligned so that each model has approximately 1.2 million trainable parameters. Under this matched parameter budget, the proposed MMPD achieves \textcolor{black}{the strongest performance among the considered neural decoders.}

\begin{figure*}[!t]
\centering

\subfloat[(384, 320) WRAN LDPC code\label{fig:ldpc-384-320}]{%
\begin{minipage}[t]{0.48\textwidth}
\centering
\includegraphics[width=\linewidth]{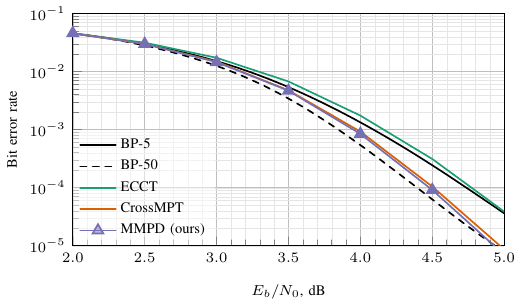}
\vspace{-14pt}
\end{minipage}%
}%
\hfill
\subfloat[(529, 440) LDPC code\label{fig:ldpc-529-440}]{%
\begin{minipage}[t]{0.48\textwidth}
\centering
\includegraphics[width=\linewidth]{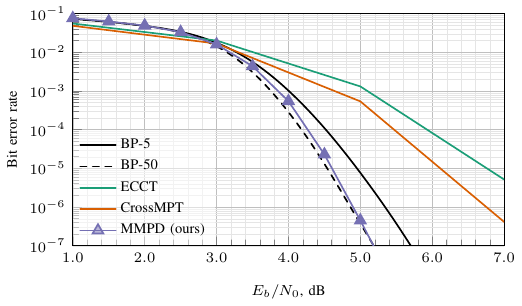}
\vspace{-14pt}
\end{minipage}%
}

\vspace{-0.5em}

\subfloat[(648, 540) IEEE 802.11n LDPC code\label{fig:ldpc-648-540}]{%
\begin{minipage}[t]{0.48\textwidth}
\centering
\includegraphics[width=\linewidth]{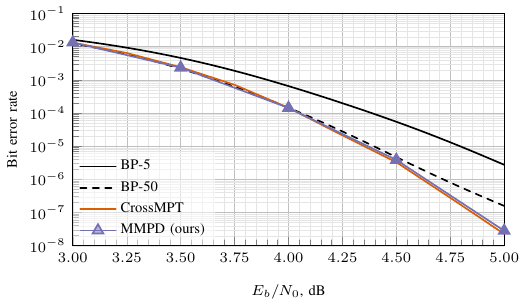}
\vspace{-14pt}
\end{minipage}%
}%
\hfill
\subfloat[(1056, 880) WiMAX LDPC code\label{fig:ldpc-1056-880}]{%
\begin{minipage}[t]{0.48\textwidth}
\centering
\includegraphics[width=\linewidth]{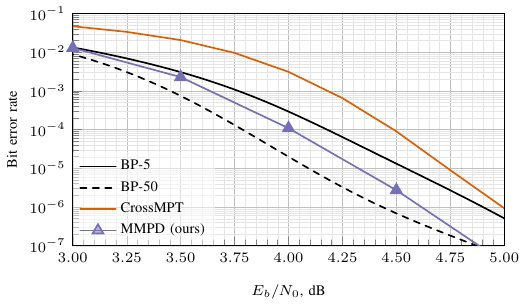}
\vspace{-14pt}
\end{minipage}%
}

\caption{BER performance for long LDPC codes.}
\label{fig:ldpc-codes-ber}
\vspace{-6pt}
\end{figure*}

In \figurename~\ref{fig:ldpc-codes-ber}, we compare the BER performance of BP with 5 and 50 iterations, ECCT, CrossMPT, and the proposed MMPD for several longer codes established as benchmarks~\cite{Park2025}, namely the $(384,320)$ wireless regional area network (WRAN) LDPC code, the $(529,440)$ LDPC code, the $(648,540)$ IEEE 802.11n LDPC code, and the $(1056,880)$ WiMAX LDPC code.

\begin{table*}[t]
\caption{Model settings and memory usage for long-code LDPC experiments}
\vspace{-20pt}
\label{tab:params-longer-ldpc}
\begin{center}
\begin{small}
\resizebox{\textwidth}{!}{
\begin{tabular}{cccccccccc}
\toprule
\multirow{2}{*}{\textbf{Codes}} & \multirow{2}{*}{\textbf{$(n,k)$}}
& \multicolumn{4}{c}{\textbf{CrossMPT}}
& \multicolumn{4}{c}{\textbf{MMPD (ours)}} \\
\cmidrule(lr){3-6} \cmidrule(lr){7-10}
&
& \textbf{Params, M}
& \textbf{Train Mem, GB}
& \textbf{Infer Mem, GB}
& \textbf{Hyperparams}
& \textbf{Params, M}
& \textbf{Train Mem, GB}
& \textbf{Infer Mem, GB}
& \textbf{Hyperparams} \\
\midrule
\multirow{4}{*}{Longer LDPC}
    & $(384,320)$  & 0.26 & 3.69  & 0.041 & T = 6, d = 32   & 0.26 & \textbf{2.35}  & \textbf{0.035} & T = 4, d = 32 \\
    & $(529,440)$  & 0.43 & 6.82  & 0.055 & T = 6, d = 32   & 0.42 & \textbf{4.18}  & \textbf{0.037} & T = 4, d = 42 \\
    & $(648,540)$  & 2.57 & 20.84 & 0.082 & T = 10, d = 128 & 2.57 & \textbf{15.05} & \textbf{0.047} & T = 8, d = 84 \\
    & $(1056,880)$ & 1.42 & 23.92 & 0.096 & T = 6, d = 32   & 1.43 & \textbf{15.83} & \textbf{0.064} & T = 6, d = 68 \\
\bottomrule
\end{tabular}
}
\end{small}
\end{center}
\vspace{-16pt}
\end{table*}

As shown in \figurename~\ref{fig:ldpc-384-320}, ECCT does not reach the performance of BP-5 on the WRAN $(384,320)$ LDPC code. In contrast, MMPD slightly outperforms CrossMPT, with its BER curve lying between BP-5 and BP-50. For the longer $(529,440)$ LDPC code in \figurename~\ref{fig:ldpc-529-440}, both ECCT and CrossMPT fall substantially short of BP-5, while MMPD approaches BP-50. \figurename~\ref{fig:ldpc-648-540} shows the results for the $(648,540)$ IEEE 802.11n LDPC code, where CrossMPT and MMPD achieve comparable performance and both outperform BP-50. Finally, for the $(1056,880)$ WiMAX LDPC code in \figurename~\ref{fig:ldpc-1056-880}, CrossMPT remains below BP-5, whereas MMPD performs between BP-5 and BP-50 and approaches BP-50 in the high-SNR regime\footnote{In this experiment, BP exhibits a high-SNR error floor, which can be attributed to its message-passing nature and code irregularity, whereas MMPD continues to improve with SNR.}, supporting the claimed scalability to longer codes.

To further assess the practical applicability of the considered neural decoders, Table~\ref{tab:params-longer-ldpc} reports the number of trainable parameters, the GPU memory footprint during a training step with a fixed batch size of 128, and the memory required for single-sample inference. All experiments were conducted using FP32 precision on a server running Ubuntu 24.04, Python 3.12, PyTorch 2.6.0, and an NVIDIA H200 GPU with driver version 580.126.09.

We next evaluate how the memory requirements of the considered neural decoders scale on a family of practical 5G LDPC codes~\cite{3gpp-decoding}. All models are evaluated with the same fixed architectural hyperparameters, such as the number of layers, hidden dimension, and number of decoding iterations, while the code length is varied. We consider the 5G LDPC codes with rate $11/12$ constructed from the BG1 base graph with $k_b=22$. By varying the lifting factor, we obtain block lengths, including punctured bits, ranging from $n=208$ to $n=9984$.

\begin{figure}[!t]
\centering
\includegraphics[width=\columnwidth]{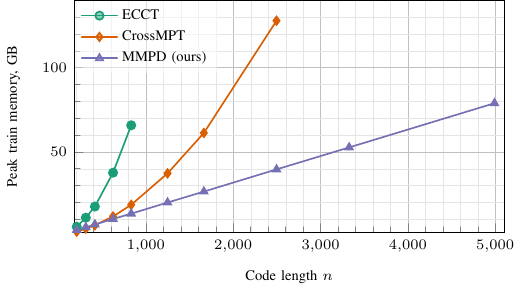}
\vspace{-22pt}
\caption{Peak GPU memory footprint during training for 5G LDPC codes, measured with a batch size of 128. The available memory is limited by the RAM size of a single GPU.}
\label{fig:ldpc-5g-memory}
\end{figure}

\begin{figure}[!t]
\centering
\includegraphics[width=\columnwidth]{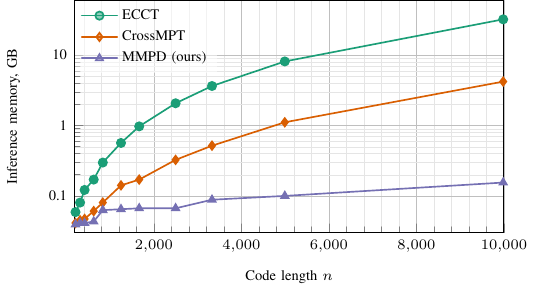}
\vspace{-22pt}
\caption{GPU memory footprint during single-sample inference for 5G LDPC codes.}
\label{fig:ldpc-5g-memory-inference}
\end{figure}

\begin{figure}[!t]
\centering
\includegraphics[width=\columnwidth]{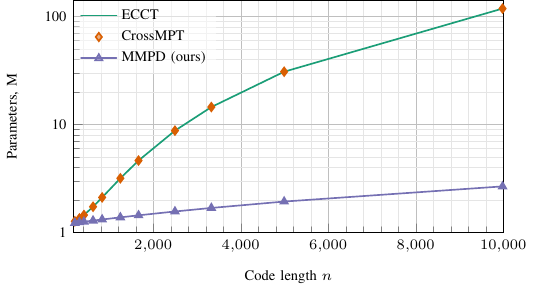}
\vspace{-22pt}
\caption{Number of trainable parameters for 5G LDPC codes}
\label{fig:ldpc-5g-params}
\vspace{-16pt}
\end{figure}

Figs.~\ref{fig:ldpc-5g-memory} and~\ref{fig:ldpc-5g-memory-inference} report the measured memory required for training with a batch size 128 and for single-sample inference, respectively. \figurename~\ref{fig:ldpc-5g-params} shows the corresponding number of trainable parameters.

The observed trends are consistent with the different interaction structures of the architectures. In attention-based decoders, storing pairwise attention maps introduces a quadratic memory term in the relevant sequence length. In MMPD, pairwise interactions are restricted to Tanner-graph edges, yielding an $\mathcal{O}(|E|)$ interaction-memory term, where $|E|$ is the number of edges induced by the parity-check matrix. For fixed model dimensions, the VN/CN states and the Mamba and FFN activations scale linearly with the corresponding stream lengths, so the inference path does not introduce a quadratic pairwise-memory term.

A similar effect is observed in the number of trainable parameters. For the considered attention-based architectures, the parameter growth is mainly caused by the output head. In particular, after the hidden dimension is reduced from $d$ to $1$, the output head maps the concatenated magnitude and syndrome representations of length $2n-k$ to an output vector of length $n$. As a result, this layer introduces a parameter contribution that scales as $\mathcal{O}(n(2n-k))$, which is quadratic in the code parameters. In contrast, the proposed MMPD uses a lightweight output head independent of the code parameters applied only to the magnitude stream, as illustrated in \figurename~\ref{fig:decoder-architecture}.

These results indicate that the quadratic growth in memory consumption and in the number of trainable parameters in the attention-based model-free decoders can restrict their applicability to practical long codes. The proposed attention-matrix-free MMPD, on the other hand, restricts pairwise interactions to the Tanner graph and therefore scales more favorably to larger block lengths.

\section{Conclusion}
\label{sec:conclusion}

This paper introduced MMPD, an attention-matrix-free syndrome-based neural decoder for binary linear codes. By combining Tanner-graph edge-restricted message aggregation with bidirectional Mamba state-space blocks, MMPD preserves structured variable-check reasoning while enabling efficient long-range information propagation.

MMPD achieves the best overall performance among the considered neural decoders on standard benchmarks and avoids the quadratic pairwise-memory term of dense attention: its graph-interaction memory scales as $\mathcal{O}(|E|)$, while the remaining Mamba and FFN terms scale linearly with the VN and CN stream lengths. This favorable scaling enables neural decoding at code lengths where attention-based decoders become memory-prohibitive. Future work includes evaluation on longer 5G LDPC and more
diverse codes, detailed complexity analysis, MMPD-oriented code design for improved performance and reduced complexity, and extensions toward foundation decoders capable of generalizing to unseen code structures.

\enlargethispage{3\baselineskip}

\end{document}